\documentclass[manuscript]{aastex}

\slugcomment{Submitted to The Astrophysical Journal}

\shorttitle{Kinematics of NGC 2818}
\shortauthors{R. V\'azquez}

\begin{document}

\title{Bubbles and Knots in the Kinematical Structure of the Bipolar Planetary Nebula NGC 2818}

\author{Roberto V\'azquez}
\affil{Instituto de Astronom\'{\i}a, Universidad Nacional Aut\'onoma de M\'exico,
Km 103 Carretera Tijuana-Ensenada, 22860 Ensenada, B. C., Mexico}
\email{vazquez@astro.unam.mx}

\begin{abstract}
High-resolution Hubble Space Telescope (HST) archive imaging and high-dispersion 
spectroscopy are used to
study the complex morphological and kinematical structure of the planetary nebula, 
\object{NGC 2818}. 
We analyze narrow band H$\alpha$, [\ion{O}{3}], [\ion{N}{2}], 
[\ion{S}{2}] and \ion{He}{2} images, addressing important morphological features.
Ground-based longslit echelle spectra were obtained crossing NGC 2818 at five different positions
to precisely determine kinematical features in the structure of the nebula. A distance of 2.5\,kpc
\citep{van95} was used to determine physical scales.
Constructing models to fit the data with modern computational tools, we find NGC\,2818 is composed by:
(1) a non-uniform bipolar structure with a semi-major axis of 0.92\,pc (75\,arcsec), possibly deformed 
by the stellar wind, (2) a 0.17\,pc (14\,arcsec) diameter central region, which is potentially the remnant 
of an equatorial enhancement, and (3) a great number of cometary knots. These knots are preferentially located inside a radius of 0.24\,pc (20\,arcsec) around the central star. The major axis of the main structure 
is oriented at $i\simeq60\degr$ with respect to the line-of-sight and at $\rm PA=+89\degr$ on the plane 
of the sky. Expansion velocities of this nebula are $V_{\rm pol}=105\,{\rm km\,s}^{-1}$ and 
$V_{\rm eq}=20\,{\rm km\,s}^{-1}$, which lead to our estimate of the kinematical age of 
$\tau_{\rm k}\simeq8,400\pm3,400$\,yr (assuming homologous expansion).  Our observations do not 
support the idea that high velocity collimated ejections are responsible for the formation of 
microstructures inside the nebula. We determine the systemic velocity of \object{NGC 2818} to be 
$V_{\rm HEL}=+26\pm2\,{\rm km\,s}^{-1}$.
\end{abstract}

\keywords{planetary nebulae: individual: NGC 2818 --- ISM: kinematics and dynamics}

\section{Introduction}
\label{intro}

\subsection{Morphology of Planetary Nebulae}
The Planetary Nebula (PN) phase occurs late in the evolution of low-mass stars  when
UV radiation from the star produces a photoionization-recombination balance in an 
expanding gaseous shell which had been expelled during the asympototic giant branch 
(AGB) phase.  Supersonic expansion velocities of 30-40 km\,s$^{-1}$ are seen in many
planetary nebulae (PNe) shells, in which the sound velocity is $\sim10$\,km\,s$^{-1}$.
To explain why the PNe do not follow a free thermal expansion (sound velocity), \citet{kwo78} 
proposed the Interacting Stellar Wind (ISW) model, whereby a denser and slower AGB wind 
interacts with a new fainter and faster stellar wind.  The existence of a fast wind around the central 
star of PN was confirmed years later by means of UV observations with the International 
Ultraviolet Explorer (IUE) satellite \citep[e.g.][]{hea78}.

Improvements in optical imaging have revealed that an understanding of the morphology of 
PNe requires more complex explanations and depends on additional physical processes. 
\citet{bal87} proposed a modification to the ISW model, the Generalized Interacting Stellar
Wind (GISW) model, which includes an equatorial density enhancement to explain elliptical
and bipolar morphologies. This model was subsequently confirmed as an appropriate
approximation and further developed by \citet{ick88} and \citet{fra94}. Since then, many other
morphological features have been discovered in deeper and better resolved observations:
open bipolar lobes \citep[called `diabolos'; e.g.][]{mon00}, multipolar structures
\citep[e.g.][]{lop98,gue98,vaz08}, rings \citep[e.g.][]{vaz02,bon03,mit06,jon10}, and
microstructures (also called Low Ionization Structures or LIS) such as ansae, knots, jets and
filaments \citep{gon01,mis09b}. In some cases, point-symmetry appears as an additional 
ingredient to the general morphology \citep[e.g.][]{gue99,sab11}. As PNe have different observed 
morphological features, many PNe cannot clearly be assigned membership to a morphological
class based on a single defining feature (round, bipolar, or elliptical).

Bipolarity has been explained since the pioneering work of \citet{bal87} as a consequence 
of an equatorial density enhancement of material formed by a mass-loaded `superwind' ejected
in the late-AGB phase. However, it is still poorly understood how this superwind is triggered and 
why it diverges from a spherical ejection. Considering these open questions, \citet{dem09} reviews
the possible shaping agents, remarking that binarity (whose definition is expanded to include
`normal' stars, brown dwarfs, and even planetary systems, as companions) is a necessary ingredient 
to explain non-spherical shapes and possibly plays a fundamental role for spherical ones. 
In particular, \citet{dem09} considers common-envelope binary interactions as an original 
shaping agent in the formation of bipolar PNe, taking up the work of \citet{sok97}.
Observations to probe binarity in central stars of PNe are non-trivial, but have been successful
in a number of cases \citep[e.g.][]{bon79,bon92,bon95,bon00,dem04,sor04,afs05} and more recently
with different techniques \citep{mis08,mis09a,fre07,mis11}.  A large fraction of PNe have a central star
which is actually a close binary system.  \citet{mis09b} note the fraction of close binary systems is
$\simeq17\pm5\,\%$, although there is evidence this number could be higher \citep{dem09}.
Currently, the number of close binary systems acting as central stars of planetary nebulae (CSPN) with 
orbital periods less than one day is $\sim$40 \citep{mis11}. 

Scenarios other than a close binary system have also been explored.
Numerical simulations demonstrate it is possible to get an equatorial density enhancement in the
superwind when stellar rotation and/or magnetic fields are included in the model \citep{gar99,gar05}.
However, other studies have found that stellar differential rotation can be drained by the magnetic
field itself \citep{sok06,nor07}, reducing the field lifetime to $\sim100$\,yr, too short to influence the
geometry of the gaseous ejecta.  More work is needed to determine the extent to which
stellar rotation and magnetic fields are important in the formation of single-star 
axisymmetrical PNe.
  
To investigate the origin of PNe morphology and to characterize the important physical
processes at work, theoretical models need to be supported with
more high quality observational data. Morphokinematic studies are a fundamental 
technique available to approach the problem \citep{kwo00}, in which high resolution images and spectra are 
used to determine the real 3-D form of the complex gaseous structures.
In this work we contribute to the study of PNe by modeling the morphokinematic
structure of \object{NGC 2818}, an object whose highly detailed images have been
published in the Hubble Heritage Project website\footnote{http://heritage.stsci.edu}.

\subsection{NGC 2818}
\label{intro2}
Previous studies of \object{NGC 2818} include the detailed work by \citet{duf84}, who analyzed 
wide band imagery, photoelectric photometry, and optical and IUE spectroscopy.  Dufour reports
the following physical parameters: $T_{\rm e}$([\ion{O}{3}])$=14\,500\pm500$\,K, 
$T_{\rm e}$([\ion{N}{2}])$=11\,500\pm800$\,K, 
$N_{\rm e}$([\ion{S}{2}])$=430\pm250\,{\rm cm}^{-3}$, and 
$N_{\rm e}$([\ion{Cl}{3}])$=1300\pm800\,{\rm cm}^{-3}$, and a heliocentric systemic velocity of 
$V_{\rm HEL}=+18.5\pm1.7\,{\rm km\,s}^{-1}$. 
\citet{sch95} obtained infrared images of \object{NGC 2818}, in which the central region was clearly
seen.  This region was more intense in the south, which was suggested to be related to shock
excitation.

\citet{phi98a} obtained narrow band H$\alpha$, [\ion{O}{3}], [\ion{N}{2}], and [\ion{S}{2}] images
from which some ratios were performed. In spite of the the low resolution of such images, the
values of the ratios along the minor axis were plotted in diagnostic diagrams to investigate the
nature of the emission. They did not directly observe clumps, but were able 
to deduce a clumpy medium formed by knots and filaments based on variations in electron density 
in the [\ion{S}{2}] images. They also proposed a possible shock excitation for the southern region
along the minor axis seen in ${\rm H}_2$ by \citet{sch95}, estimating planar shocks with velocities 
of $V_{\rm s}\ge110$\,km\,s$^{-1}$.
Better resolution H$\alpha$ and [\ion{O}{3}] images from \citet{gor99} 
partially show the complex internal structure proposed by \citet{phi98a}. A recent study in the 
{\it Spitzer} mid-infrared (MIR) bands (3.6, 4.5, 5.8, and 8\,$\mu$m) by \citet{phi10} shows
gradients in the emission ratios.  The observed gradients indicate an increasing 
importance of emission from photodissociation regions over emission from photoionization. 
Other contributions to the observed MIR emission include emission from polycyclic armatic hydrocarbons (PAHs) and effects of shocked regions.

Finally, the study of \object{NGC 2818} is of particular interest because it has been the topic 
of a series of articles in which its membership to a stellar open cluster 
is debated. This is an important issue given that 
distance is a parameter difficult to estimate for PNe, but relatively easy to estimate for stellar 
clusters. In addition, there are only a few cases of PNe physically related to stellar clusters 
\citep[e.g.][]{par11}. Addressing this issue is outside the scope of this paper, but a detailed study
to address this question is in preparation.

\section{Observations and results}

\subsection{Imaging}
\label{imaging}

\begin{figure*}
\centering
\includegraphics[width=\linewidth]{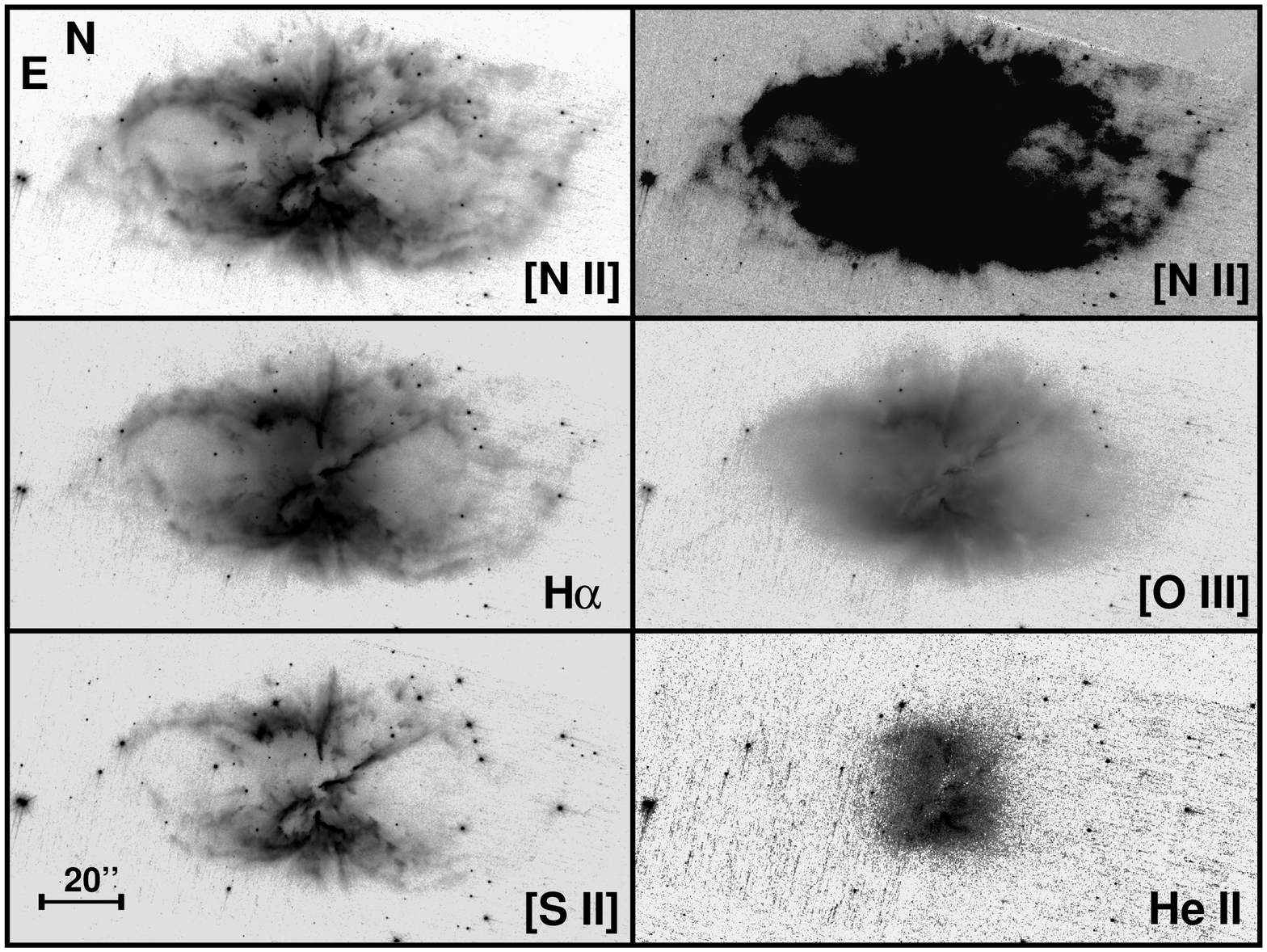}
\caption{HST-WFPC2 narrow band images. Upper panels: [\ion{N}{2}]$\lambda6583$ (filter F658N)
with two different contrasts; central panels: H$\alpha$ (left; filter F656N) and [\ion{O}{3}]$\lambda5007$
(right; filter F502N);
lower panels: [\ion{S}{2}]$\lambda6731$ (left; filter F673N) and \ion{He}{2}\,$\lambda4686$ (right; filter
F469N). Spatial scale is indicated. Sky orientation (north up, east left) is the same in all subsequent figures.}
\label{mosaico}
\end{figure*}

A series of Hubble Space Telescope (HST) narrow band images, taken with the Wide Field Planetary 
Camera 2 (WFPC-2), were retrieved from the MAST Archive\footnote{This study is based on observations 
made with the NASA/ESA HST, obtained from the data archive at the Space Telescope Science Institute 
(STScI). STScI is operated by the Association of Universities for Research in Astronomy, Inc. under NASA 
contract NAS 5-26555.} to study the apparent morphology of \object{NGC 2818} (proposal ID: 6119; PI: H. E. 
Bond; Date of observation: 1995 August 26). We show these images in Figure~\ref{mosaico}, presenting 
images with different contrasts for [\ion{N}{2}]$\lambda6583$ emission (filter F658N) in the two upper panels, 
H$\alpha$ (left) and [\ion{O}{3}]$\lambda5007$ (right) emission lines (filters F656N and F502N) in the central 
panels, and [\ion{S}{2}]$\lambda6731$ (left) and \ion{He}{2}\,$\lambda4686$ (right) emission lines (filters 
F673N and F469N) in the bottom panels. Three {\it drizzled} images were combined for each filter in order to 
remove cosmic rays and bad 
pixels according to standard procedures in HST reduction manuals. 

The appearance of \object{NGC 2818} is slightly different when observed in different filters,
as expected due to ionization stratification around the central star of a PN.
In Fig.~\ref{mosaico}, the basic structure appears to be composed of two bipolar lobes,
separated in the center by a slightly pinched waist. The enhanced brightness in the cusps of the waist is 
similar to the limb-brightness of a denser squeezed ring-like structure. We will refer to this area around 
the central star, clearly seen as the unique emission region in \ion{He}{2}, as `the central region'. 

Both the bipolar lobes and the central region are visible in all panels, with the exception 
of  \ion{He}{2}, in which only the central region is visible.  We note that the [\ion{O}{3}] emission 
appears to be a smooth distribution of gas, filling the entire nebula.
The length of the major axis of the nebula projected on the plane of the sky is $\simeq101$\,arcsec, 
whereas the minor axis is 31\,arcsec, with the smallest separation of 14\,arcsec at the cusps of the waist.
The whole nebula can be contained in an ellipse of $126\times62$\,arcsec.

In addition to the main structure, a series of microstructures (cometary knots) are especially visible
in the [\ion{N}{2}] image, but can also be seen in the [\ion{S}{2}] and H$\alpha$ images
(Fig.~\ref{mosaico}, left panels). The tails of these cometary knots are primarily directed outwards
from the geometrical center of the nebula. One of these microstructures, which is very prominent
in the central region of the nebula, appears as an asymmetrical and broken `big tail', almost aligned
SE-NW. The upper-right panel of Fig.~\ref{mosaico} shows a low contrast [\ion{N}{2}] image in order
to highlight the surroundings of the main structure. At the edges, several
tails are also detected that can be traced outwards from the nebula.

We made ratio maps using the HST images (Fig.~\ref{mosaico}) to emphasize 
different morphological features in \object{NGC 2818}. Fig.~\ref{cocientes}
shows a mosaic of four image ratios. The upper-left panel shows the [\ion{N}{2}]/H$\alpha$ 
ratio in which the low-ionization regions (white) are better defined than in
the original images. The image ratio [\ion{O}{3}]/H$\alpha$ is sensitive to the
variation of ionization/excitation and chemical abundances \citep[see][]{cor96}
and can also be used to look for signatures of collimated outflows \citep{med07}.
Given the nature of the involved ions, the [\ion{S}{2}]/H$\alpha$ image ratio is 
sensitive to shocked regions whereas the \ion{He}{2}/H$\alpha$ image ratio 
shows the very high ionization regions. 

\begin{figure*}
\centering
\includegraphics[angle=270, width=0.8\linewidth]{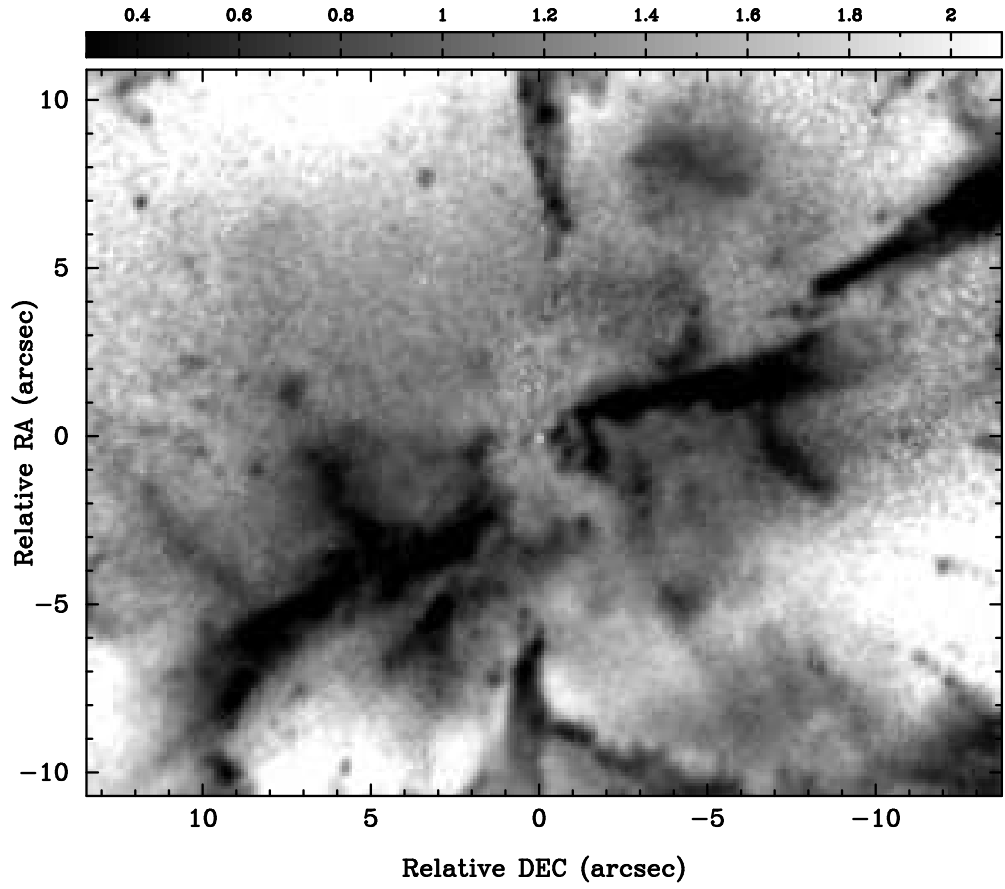}
\caption{Detailed view of the central region in the [\ion{O}{3}]/H$\alpha$ image ratio
presented in Fig.~1. The central white dot corresponds to the central star of NGC 2818.}
\label{zoom}
\end{figure*}

The location of the central star of \object{NGC 2818} can be easily found by looking for a hot
star in a blue-to-red image ratio. Fig.~\ref{zoom} shows an enlargement of the central
region of the nebula in the [\ion{O}{3}]/H$\alpha$ image in which the central star appears 
as  a central white dot. The location of the central star is in agreement with the extrapolation
of the radial tails of many knots (not shown), as expected if they originate from the identified
central star position.

\subsection{Longslit high-dispersion spectroscopy}

\begin{figure*}
\centering
\includegraphics[width=\linewidth]{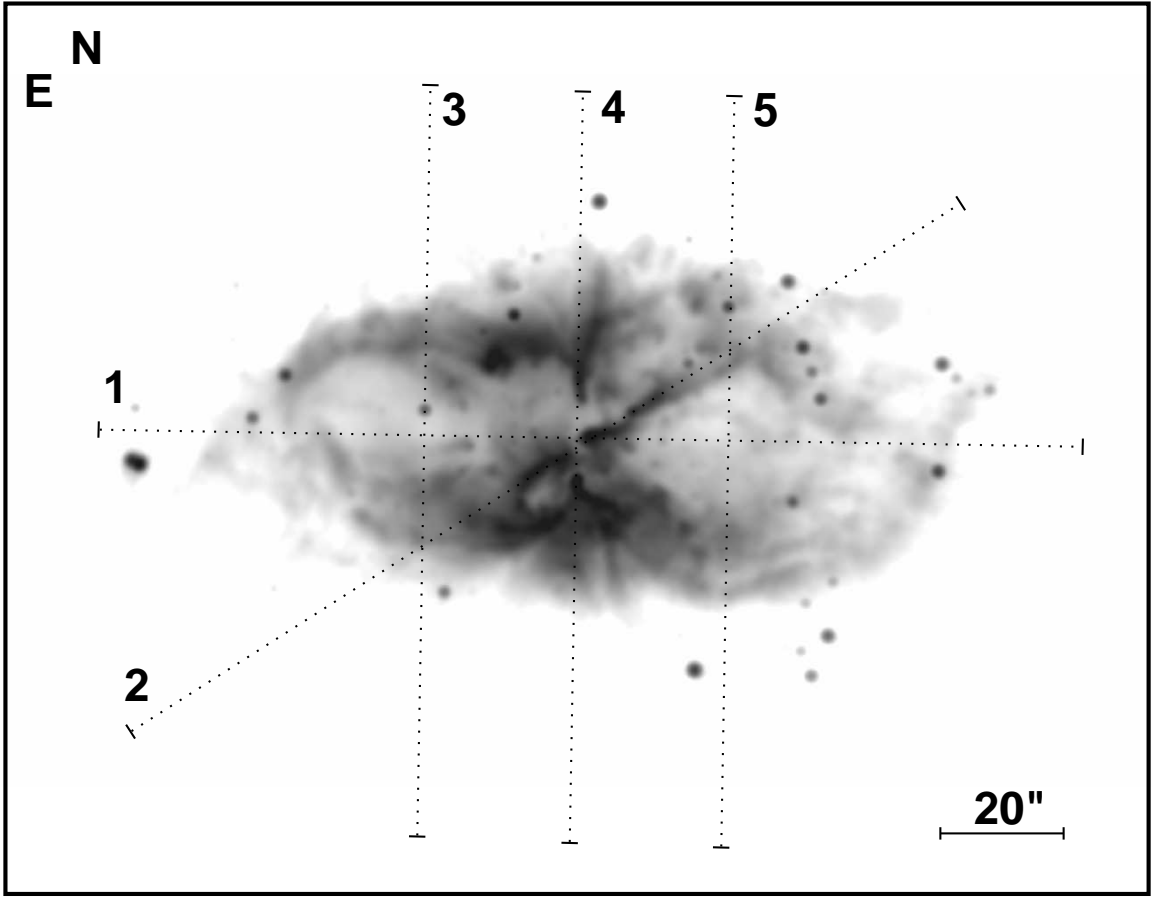}
\caption{Slit positions used in the high dispersion spectroscopy
superimposed on an [\ion{N}{2}] HST image.
PAs=$+89\degr$ (1), $+123\degr$ (2), and $-1\degr$ (3, 4, and 5). The image was
gaussian filtered to reproduce the spatial resolution observed in the spectra.}
\label{slits}
\end{figure*}

High-dispersion optical spectra were obtained with the Manchester Echelle
Spectrometer \citep[MES;][]{mea03} and the 2.1\,m (f/7.5) telescope at the 
OAN-SPM\footnote{The Observatorio Astron\'omico Nacional at the Sierra de 
San Pedro M\'artir (OAN-SPM) is operated by the Instituto de Astronom\'{\i}a 
of the Universidad Nacional Aut\'onoma de M\' exico} observatory during 2009 
February 4$-$6. A SITe CCD with $1024\times1024$ pixels was used as detector. 
The slit length is $6\farcm5$ and the width was set to 150-$\mu$m (2\arcsec).
A $2\times2$ binning was used, leading to a spatial scale of 0\farcs6\,pixel$^{-1}$ 
and a spectral scale of 0.1\,{\AA}\,pixel$^{-1}$. This spectrograph has no cross
dispersion, consequently a  $\Delta\lambda=90$\,{\AA} bandwidth filter was used 
to isolate the 87$^{\rm th}$ order covering the spectral range which includes
H$\alpha$ and [\ion{N}{2}]$\lambda6583$ emission lines. The spectra were
wavelength calibrated with a Th-Ar arc lamp to an accuracy of $\pm1$\,km\,s$^{-1}$. 
The FWHM of the arc lamp emission lines was measured to be $\simeq12$\,km\,s$^{-1}$. 
Spectra were obtained by positioning the slit across five different regions of the nebula,
as shown in Figure~\ref{slits}. The position angles (PAs) for these slits are $+89\degr$
(slit 1), $+123\degr$ (slit 2), and $-1\degr$ (slits 3, 4, and 5). Exposure time for all the 
slits was 1800\,s.  Seeing was $\sim$2 arcsec during observations.

Fig.~\ref{pvs} shows the Position-Velocity (PV) maps for H$\alpha$ (upper panels)
and [\ion{N}{2}]\,$\lambda6583$ (lower panels) emission lines. Each column is labeled 
with the corresponding slit number. We note that the \ion{He}{2}$\lambda6560$ line 
is also observed in the PV maps for slits 1, 2, and 4, in agreement with the highest
excitation location shown in the \ion{He}{2} image (see lower-right panel in Fig.~\ref{mosaico} ).
The relative position is measured with respect to the position of the central star 
(for slits 1, 2, and 4) and to the major axis defined by slit 1 (for slits 3 and 5). Radial 
velocity is relative to the systemic velocity $V_{\rm HEL}=+26\pm2$\,km\,s$^{-1}$ 
($V_{\rm LSR}=+11\pm2$\,km\,s$^{-1}$). This value was deduced as the mean
velocity of the two components of the splitting of the spectral lines, taken from the
slit region crossing the central star (on slits 1, 2, and 4). This method
only can be done if the splitting of the spectral lines can be measured; that is, when we
have high-resolution spectra. It has better accuracy than only taking the maximum
of a broad unique emission line, as can be seen in low-resolution spectra. Observing
without enough spectral resolution to see the splitting can result in a wrong estimate
for the systemic velocity if, as in this case, the components of the splitting have different
brightness. Without high resolution spectroscopy, the non-resolved splitting is seen as 
a broad emission line with its maximum being biased towards the wavelength of the
brightest component. Previous determinations of the systemic velocity of \object{NGC 2818}
have been been affected by low resolution spectra 
\citep[e.g., $V_{\rm hel}=18.5\pm1.7$\,km\,s$^{-1}$,][]{duf84}.
In this paper, we present the first systemic velocity determination from the splitting method,
utilizing high resolution spectroscopy.

\section{Discussion}
\label{discusion}

\subsection{The morpho-kinematic structure of NGC 2818}
\label{structure}

\begin{figure*}
\centering
\includegraphics[angle=270, width=\linewidth]{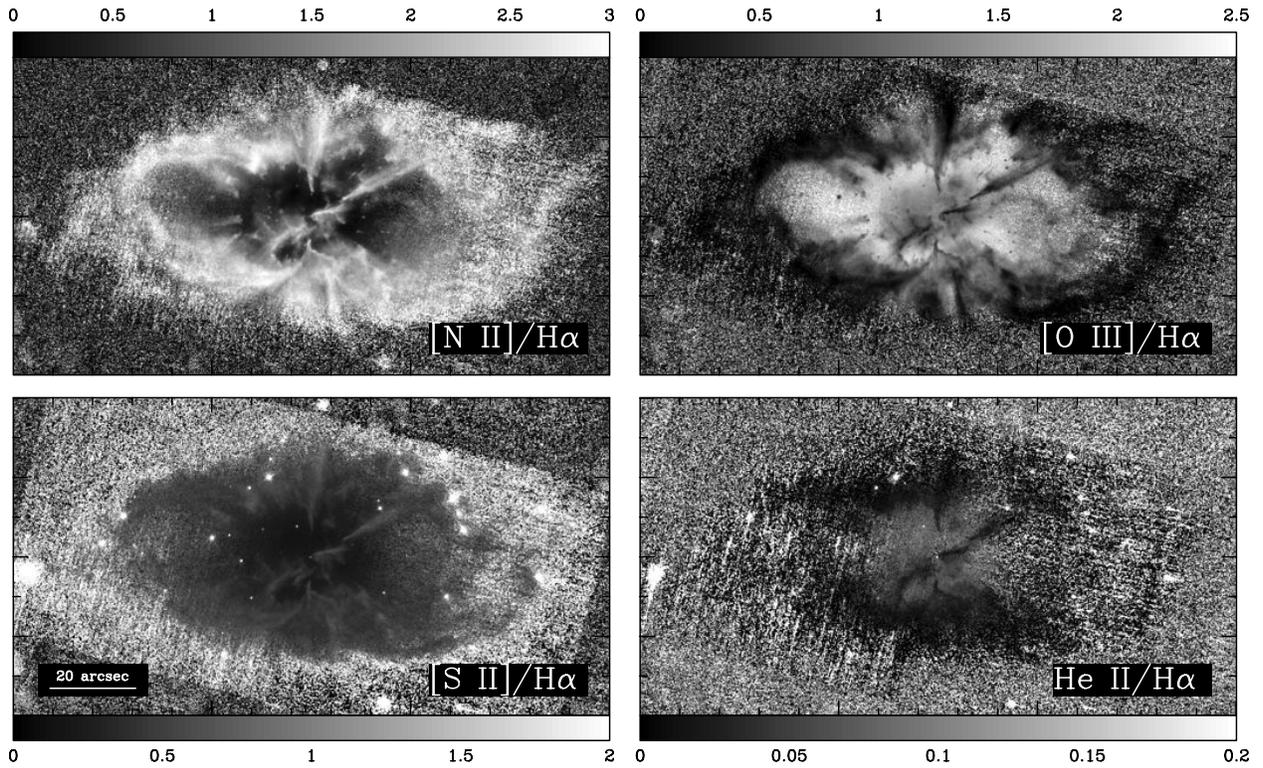}
\caption{HST-WFPC2 narrowband image ratios of NGC\,2818. White corresponds to large values of each ratio, indicated by the gray-scale bar.
Upper left: [N\,{\sc{ii}}]/H$\alpha$. Upper right: [O\,{\sc{iii}}]/H$\alpha$. Lower left: [S\,{\sc{ii}}]/H$\alpha$.
Lower right: He\,{\sc{ii}}/H$\alpha$.}
\label{cocientes}
\end{figure*}

\begin{figure*}
\centering
\includegraphics[width=0.8\linewidth]{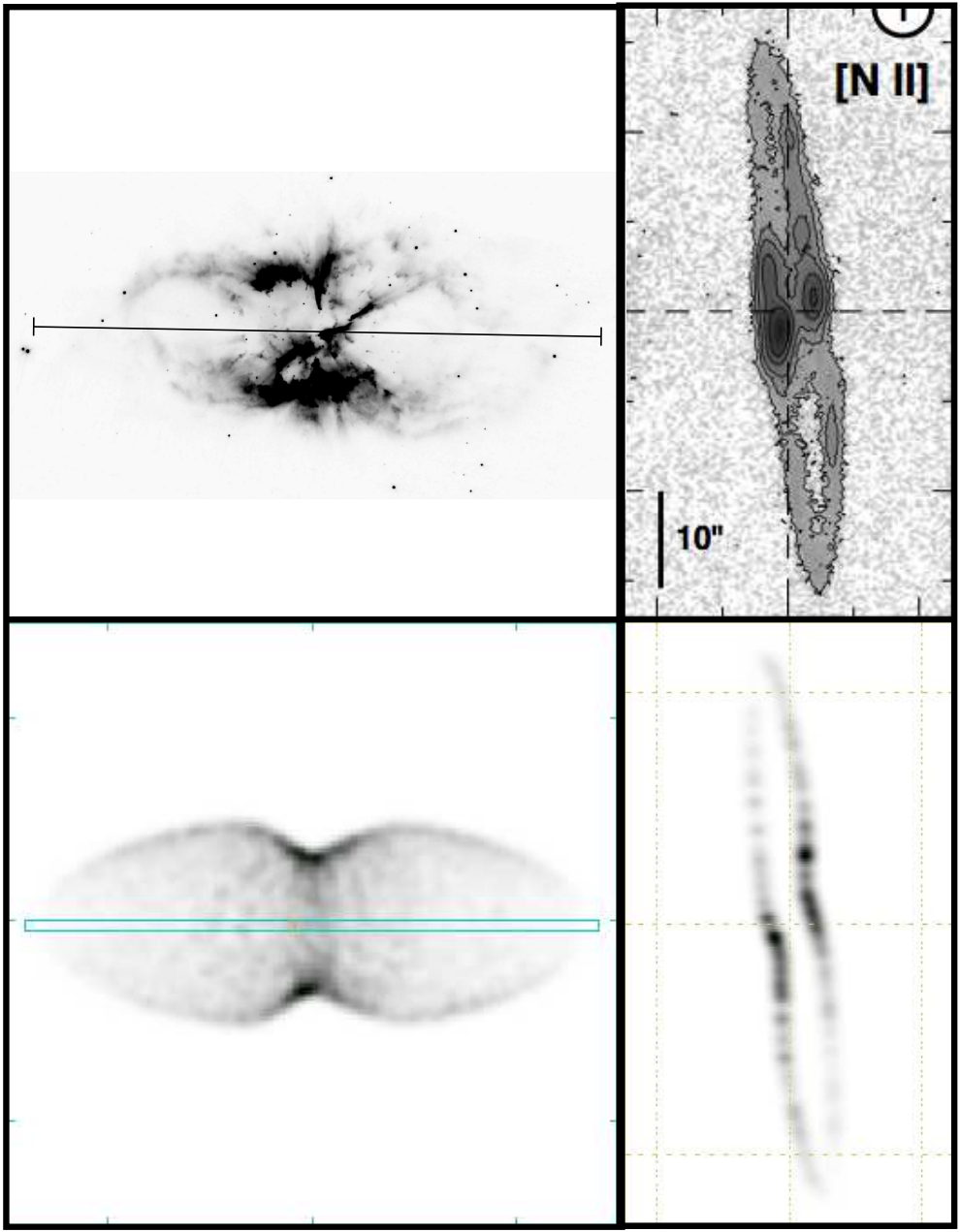}
\caption{Comparison of observed data and the 3D model obtained from {\sc shape}. Left: Observed
[N\,{\sc ii}] image (upper panel) and the {\sc shape} image (lower panel). Right: Observed PV map of
the [N\,{\sc ii}]\,$\lambda6583$ line at slit position 1 (upper panel) and the
corresponding {\sc shape} synthetic PV map (lower panel).}
\label{shape}
\end{figure*}

As described above (Sec.~\ref{imaging}), our inspection of the unprecedented high resolution HST images 
indicates that \object{NGC 2818} is a morphologically rich PN with minor bubbles, asymmetrical outflows, 
and cometary knots inside its main structure (Figs.~\ref{mosaico} and \ref{cocientes}). In a first attempt to 
study the morphology and kinematics of \object{NGC 2818}, we have constructed some simple models using 
the interactive software {\sc shape} \citep{ste11}, a computational tool for the morpho-kinematic modeling 
and reconstruction of astrophysical objects. The internal library of {\sc shape} includes 
basic structures such as spheres, cylinders, and tori, which can be modified by applying different parameters 
and/or functions (``modifiers''). Each structure is composed of a grid in which a certain number of points are 
distributed. The user defines the number of points for each structure and the volume or surface distribution. 
Physical parameters, such as density and velocity, can be assigned to the points either as an analytical 
function or interactively. Several recent papers have used {\sc shape} to get the structure of PNe 
\citep[e.g.][]{mis11,hsi10,jon10,con10}. In our particular case, we obtained a 
final nebula model which fits the main structure of \object{NGC 2818} for both imaging and spectroscopy 
(Fig.~\ref{shape}). We find \object{NGC 2818} can be modeled as a bipolar structure with major and minor 
axis length of $1.8\times0.3$\,pc \citep[$150\times28$\,arcsec at a distance of 2.5\,kpc, following][]{van95},
with an inclination angle of the major axis equal to 60{\degr} with respect to the line of sight (eastern lobe is 
blueshifted, western lobe is redshifted).  The polar and equatorial velocities are
$V_{\rm pol}=105\,{\rm km\,s}^{-1}$ and $V_{\rm eq}=20\,{\rm km\,s}^{-1}$, respectively, assuming 
homologous velocity expansion ($V\propto r$). The shell has a maximum thickness of 0.1\,pc 
($\simeq9$\,arcsec) becoming thinner towards the poles. Using these velocities and dimensions, we 
have estimated a kinematical age of $\tau_{\rm k}\simeq8,400\pm3,400$\,yr, which is similar to the age of 
other PNe with cometary knots in their morphology \citep[see][]{ode02}.

The relatively high uncertainty in the kinematical age estimate merits a more detailed explanation. 
Distance estimates for PNe remain an important issue, given the very few cases in which we can get
a `direct' measurement (e.g. via stellar parallax).  We have avoided distance estimates for NGC\,2818
based on membership of this object to the stellar cluster mentioned in Sec.~\ref{intro2}, and instead have 
used statistical values. There are some estimates for the distance to this object ranging from 1.79\,kpc
\citep{phi04} to 3.05\,kpc \citep{zha95}. \citet{van95} give a value of 2.5\,kpc, which is very close to
mean value of the distance range. An uncertainty of $\sim$40\% is obtained utilizing the
statistical estimates (which covers all the range of distance estimates). Thus, the propagation of this 
high uncertainty has effects in the same proportion over the physical size ($1.8\pm0.7$\,pc in the major axis) 
and the kinematical age ($8,400\pm3,400$\,yr).  Having better distance estimates will improve the accuracy 
on these two parameters, and is the focus of future work.

\begin{figure*}
\centering
\includegraphics[width=0.8\linewidth]{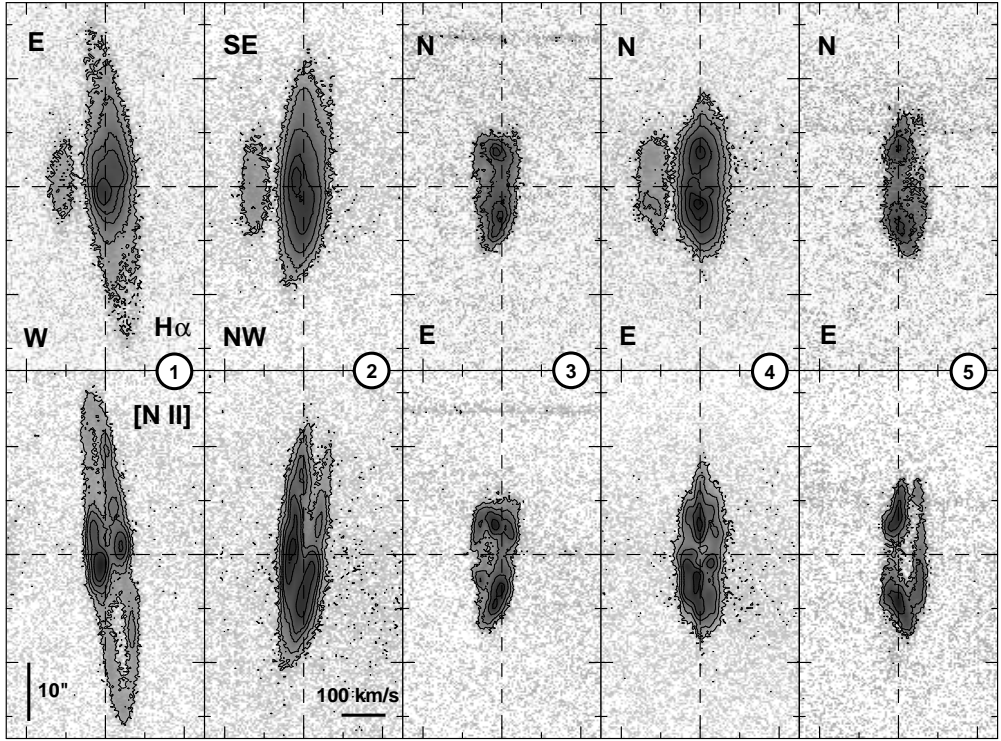}
\caption{Position-Velocity maps of NGC\,2818 for
slits positions 1 to 5. Upper panels correspond to H$\alpha$,
and lower panels to [N\,{\sc ii}]\,$\lambda$6583.
The slit orientation indicated in the upper panels is valid for the two maps in each column. The horizontal
dashed lines represent the position of the central star for slits 1, 2, and 4, and the location where slits
cross the major axis for slits 4 and 5. The horizontal dashed line corresponds to the systemic radial
velocity ($V_{\rm HEL}=+26\pm2$\,km\,s$^{-1}$).}
\label{pvs}
\end{figure*}

Because of their faintness, kinematics from the microstructures cannot be determined as they are 
not seen in our spectra, with the exception of the `big tail' mentioned in Sec.~\ref{imaging}. A careful 
analysis of images in Figs.~\ref{mosaico}, \ref{cocientes}, and \ref{zoom}, as well as the PV map in 
Fig.~\ref{pvs} (panel 2, lower) indicates that the `big tail', as well as the other similar structures, have 
the same nature as the rest of the smaller cometary knots, being possibly photoevaporated, blown,
and ionized by the action of the wind and radiation from the central star over the dense clumps that 
remain in its surroundings. In the `big tail' a collimated ejection is unlikely
as the two NW pieces do not share the same kinematic behavior, as would be expected in the 
case of a collimated ejection (see Fig.~\ref{pvs}, panel 2, lower).
If the `big tail' is not taken into account, it is remarkable to note that the small cometary knots have 
no apparent preferred direction, whereas the denser and larger features are related to the central region. 
The latter seems to be similar to the case of the Helix nebula \citep{mea98}, which has been 
proposed to have a knotty torus whose axis is seen at $37\degr$ with respect to the line of sight.

Finally, the cometary knot antipodes in the upper-left side of the central region of \object{NGC 2818} 
appear to point towards a kind of `local' bubble center instead of the central star. One possible 
interpretation is that the peculiar kinematics of several bubbles inside the nebula play an important 
role in the production of tails. We think that this idea deserves more attention in theoretical studies 
and is beyond the scope of the present work.

\subsection{The evolution of NGC 2818}

In this section we consider the line ratio images and what these ratios can tell us about the evolution 
of \object{NGC 2818}.  First, we examine the image ratio [\ion{O}{3}]/H$\alpha$. This ratio is sensitive 
to the variation of ionization/excitation and chemical abundances \citep[see][]{cor96} and can also be 
used to look for signatures of collimated outflows \citep{med07}. According to \citet{med07}, PNe with 
[\ion{O}{3}]/H$\alpha$ ratios enhanced in an outer layer (or ``skin''), are related to both, fast collimated 
outflows and multiple-shells, the latter may even have FLIERs \citep[Fast Low-Ionization Emission 
Regions;][]{bal87}. However, as seen in Fig.~\ref{cocientes} (upper-right panel), \object{NGC 2818} 
has a [\ion{O}{3}]/H$\alpha$ ratio diminished (Medina's type C), which could correspond to either young 
or evolved PNe, but not related to jets, shocks, or multiple shells. It is important to note that no correction
for extinction was applied. In particular, this ratio image could be more affected by interstellar
extinction (or internal extinction, including internal small-scale variatons) than the other ratios, given the 
relative large difference between the wavelengths of [\ion{O}{3}] and H$\alpha$.

The [\ion{S}{2}]/H$\alpha$ image ratio (Fig.~\ref{cocientes}, lower-left
panel) is sensitive to the shocked regions and is enhanced in the most noticeable
tails seen in [\ion{N}{2}]. These could indicate either that these tails are shocked gas or that gas
in the tails is being shocked by the stellar wind. Following the homologous velocity assumption,
we expect no more than 28 km\,s$^{-1}$ in stellar wind velocity at the distances in which the
cometary knots are found, as they are preferentially located at a radius of 20\,arcsec around
the central star (0.24 pc, assuming a distance of 2.5\,kpc). However, these low velocity shocks
could be enough to give rise to the
[\ion{S}{2}]/H$\alpha$ ratio in a detectable way \citep[see][and references therein]{phi98b}.
The same ratio also appears to exist in the ring zone. This could be in agreement with
the results of \cite{sch95} and \citet{phi98a} who detect emission possibly related to shocked
gas. It is also remarkable that based on the imaging in their study, \citet{phi98a} propose velocities 
of 110\,km\,s$^{-1}$ for the farthest regions of the nebula. Based on kinematics, we
also estimate polar velocities around this value.

The \ion{He}{2}/H$\alpha$ image ratio shows the very high ionization regions. 
As expected, the highest ionized regions in \object{NGC 2818} are enclosed inside the main
structure of the nebula in the regions surrounding the central star. This can also be seen in the
the [\ion{O}{3}] and \ion{He}{2} images. In previous sections we defined the `central region' 
based on the \ion{He}{2} emission, whose appearance reminds one of the remnant of an
equatorial structure. A possible interpretation of the data could be that \object{NGC 2818} is 
a bipolar PN whose equatorial density enhancement can be traced by the cusps aligned with the 
central star and the brightness of the central region. The distorted internal structure and the 
presence of multiple bubbles and non-uniform microstructures can be seen as part of a pre-existent
clumpy medium in the surroundings. Although a definite model for the formation of cometary knots
has not been generally accepted, hydrodynamical instabilities appear to adequately explain them 
\citep[e.g.][]{cap06}. Such instabilities could be formed in the beginning of the proto-PN phase and 
appear as cometary knots once the bipolar outflow is free. Trying to understand the energetics 
involved in the whole process of formation and evolution of this nebula is an interesting way to 
address the study of the synergy between macro- and microstructures in the morphology of 
PNe.

\section{Conclusions}

We have conducted an observational study of \object{NGC 2818} using archival high resolution 
space-based imaging and high dispersion Earth-based longslit spectroscopy. We derive a systemic 
velocity of $V_{\rm HEL}=+26\pm2\,{\rm km\,s}^{-1}$. The computational software {\sc shape} was used 
to construct a model to fit our data and improve our interpretation of the overall morphokinematic 
structure of \object{NGC 2818}.  Our model is consistent with a bipolar nebula with a semi-major axis
of 0.92\,pc, possibly deformed by the stellar wind, a 0.17\,pc diameter central region that is a 
potential remnant of an equatorial enhancement, and a great number of cometary knots preferentially
located inside a radius of 0.24\,pc around the central star. The major axis of the main structure is
oriented at $i\simeq60\degr$ with respect to the line-of-sight and at $\rm PA=+89\degr$ on the plane 
of the sky. Deprojected velocities correspond to an expansion of $V_{\rm pol}=105\,{\rm km\,s}^{-1}$ and 
$V_{\rm eq}=20\,{\rm km\,s}^{-1}$, which lead to an estimate of the kinematical age of 
$\tau_k\simeq8,400\pm3,400$\,yr. Cometary knots could be related to hydrodynamical instabilities
formed before the bipolar outflow started.

\acknowledgments

This project was supported by UNAM-PAPIIT grants IN111903 and IN109509.
The author is grateful to the anonymous referee for his valuable comments and to 
his research group ``Planetosos'' for its support and fruitful
discussions, in particular to Mr. Paco Beretta for calling my attention to this
object. W. Steffen and L. Guit\'errez provided {\sc shape} and HST
advising, respectively. The author also
acknowledges the OAN staff at San Pedro M\'artir, with special mention to
Mr. Gustavo Melgoza-Kennedy (aka ``Tiky'') for support during observations.
The author is deeply grateful to Dr. Ashley Zauderer for a kind
and careful revision of the manuscript. Finally, the author would like to 
dedicate this paper in memory of Prof. Yolanda G\'omez, who was an 
active researcher in the field of Planetary Nebulae and an indefatigable 
impeller of science education and outreach, and who recently passed away.

{\it Facilities:} \facility{HST (WFPC2)}, \facility{SPM (MEZCAL)}.

\clearpage

\end{document}